\documentclass[aps,pra,groupedaddress,superscriptaddress,twocolumn,showpacs]{revtex4-1}

\usepackage[utf8x]{inputenc}
\usepackage[usenames,dvipsnames]{color}
\usepackage{bbm} 

\usepackage{amsfonts,amsmath,amssymb,stmaryrd}

\usepackage{graphicx}

\usepackage{subfigure}  % use for side-by-side figures

\usepackage{bbm} 

\usepackage{hyperref}

\usepackage{bbm}

\usepackage{epsfig}

\usepackage{mathrsfs}

\usepackage{verbatim}

\usepackage{ulem}	% underline

\renewcommand{\l}{\left(}
\renewcommand{\r}{\right)}
\newcommand{\bra}[1]{\langle#1|}
\newcommand{\ket}[1]{|#1\rangle}

\renewcommand{\H}{\hat{\mathcal{H}}}

\renewcommand{\a}{\hat{a}}
\newcommand{\n}{\hat{n}}

\newcommand{\ad}{\hat{a}^\dagger}
\newcommand{\bd}{\hat{b}^\dagger}

\renewcommand{\L}{\text{L}}
\newcommand{\R}{\text{R}}

\newcommand{\hc}{\text{h.c.}}

\usepackage{array}

\usepackage{cancel,ifthen}

\newcommand{\Komment}[2][NoInPuT]{\ifthenelse{\equal{#1}{NoInPuT}}{}{{\color{red}\sout{#1}}} {\color{blue} #2}}

\newcommand{\KommentM}[2][NoInPuT]{\ifthenelse{\equal{#1}{NoInPuT}}{}{{\color{Fuchsia}\sout{#1}}} {\color{Cerulean} #2}}

\usepackage{bm}	% bold in math mode
\renewcommand{\vec}[1]{\bm{#1}}

\begin{document}
\normalem	% changes \emph back to normal after introducing ulem package.

\title{Realization of Fractional Chern Insulators in the Thin-Torus-Limit\\ with Ultracold Bosons}

\author{Fabian Grusdt}
\affiliation{Department of Physics and Research Center OPTIMAS, University of Kaiserslautern, Germany}
\affiliation{Graduate School Materials Science in Mainz, Gottlieb-Daimler-Strasse 47, 67663 Kaiserslautern, Germany}

\author{Michael H\"oning}
\affiliation{Department of Physics and Research Center OPTIMAS, University of Kaiserslautern, Germany}


\begin{abstract}
Topological states of interacting many-body systems are at the focus of current research due to the exotic properties of their elementary excitations. In this paper we suggest a realistic experimental setup for the realization of a simple version of such a phase. We show how delta-interacting bosons hopping on the links of a one-dimensional (1D) ladder can be used to simulate the thin-torus-limit of the two-dimensional (2D) Hofstadter-Hubbard model at one-quarter magnetic flux per plaquette. Bosons can be confined to ladders by optical superlattices, and synthetic magnetic fields can be realized by far off-resonant Raman beams. We show that twisted boundary conditions can be implemented, enabling the realization of a fractionally quantized Thouless pump. Using numerical density-matrix-renormalization-group (DMRG) calculations we show that the groundstate of our model is an incompressible symmetry-protected topological charge density wave (CDW) phase at average filling $\rho=1/8$ per lattice site, related to the $1/2$ Laughlin-type state of the corresponding 2D model.
\end{abstract}

\date{\today}

\maketitle

%%%%%%%%%%%%%%%%%%%%%%%%%%%%%%%%%%%%%%%%%%%%%%%%%%%%%
%%%%%%%%%%%%%%%%%%%%%%%%%%%%%%%%%%%%%%%%%%%%%%%%%%%%%
\section{Introduction}
%%%%%%%%%%%%%%%%%%%%%%%%%%%%%%%%%%%%%%%%%%%%%%%%%%%%%
%%%%%%%%%%%%%%%%%%%%%%%%%%%%%%%%%%%%%%%%%%%%%%%%%%%%%
For a long time it was believed that distinct phases of matter can be classified entirely by the concept of spontaneous symmetry breaking, which is formulated mathematically in the Ginzburg-Landau theory. With the discovery of the quantum Hall effect \cite{Vonklitzing1980,Thouless1982} it has become clear that this classification has to be extended by the inclusion of topological orders \cite{Wen1995}. More recently, the discovery of topological insulators \cite{Kane2005,Bernevig2006a,Bernevig2006,Koenig2007,Hsieh2008} and subsequent theoretical analysis \cite{Ryu2010,Chen2011b,Chen2011a,Chen2013} have revealed that the class of topologically ordered states contains a large number of phases, at least some of which have concrete physical implementations and can be observed in experiments. 

When the quantum Hall effects were discovered \cite{Vonklitzing1980,Tsui1982}, solid state systems were the most promising candidates to search for new phases of matter. Nowadays ultracold atom experiments constitute a promising alternative platform for a systematic search for novel states of matter \cite{Bloch2008}. Their large length scales make a number of new measurement methods available, and the cleanness allows for long coherence times. These ingredients have enabled the observation of the superfluid to Mott insulator transition in the Bose Hubbard model \cite{Greiner2002a}, which can be understood from the Ginzburg-Landau paradigm. Over the last century, great effort has been devoted to the observation of \emph{topological} phase transitions in such systems, going beyond the Ginzburg-Landau paradigm. In particular systems with synthetic magnetic fields were studied \cite{Wilkin1998,Cooper1999,Regnault2003,Jaksch2003,Ruseckas2005,Juzeliunas2006,Lin2009,Dalibard2011,Cooper2011a,Kolovsky2011a,Aidelsburger2011,Aidelsburger2013,Miyake2013}, where analogues of the quantum Hall effects were predicted to be observable. Recently this goal has been achieved and topological phase transitions where observed in systems of essentially non-interacting ultracold quantum gases \cite{Jotzu2014,Aidelsburger2014}. While all possible topological phases of non-interacting fermions in arbitrary dimensions have been classified \cite{Ryu2010}, it is established that interactions can enrich the number of possible topological phases enormously, see e.g. \cite{Chen2013}. The resulting phases attract much attention because they can support exotic excitations with fractional charge and statistics \cite{AROVAS1984,Halperin1984,MOORE1991,Bonderson2006,Laughlin1983}, which have possible applications for topological quantum computation \cite{Kitaev2003,Nayak2008}.

%%%%%%%%%%%%%%%%%%%%%%%%%%%%%%%%%%%%%%%%%%%%%%%%%%%%%
\begin{figure}[b]
\centering
\epsfig{file=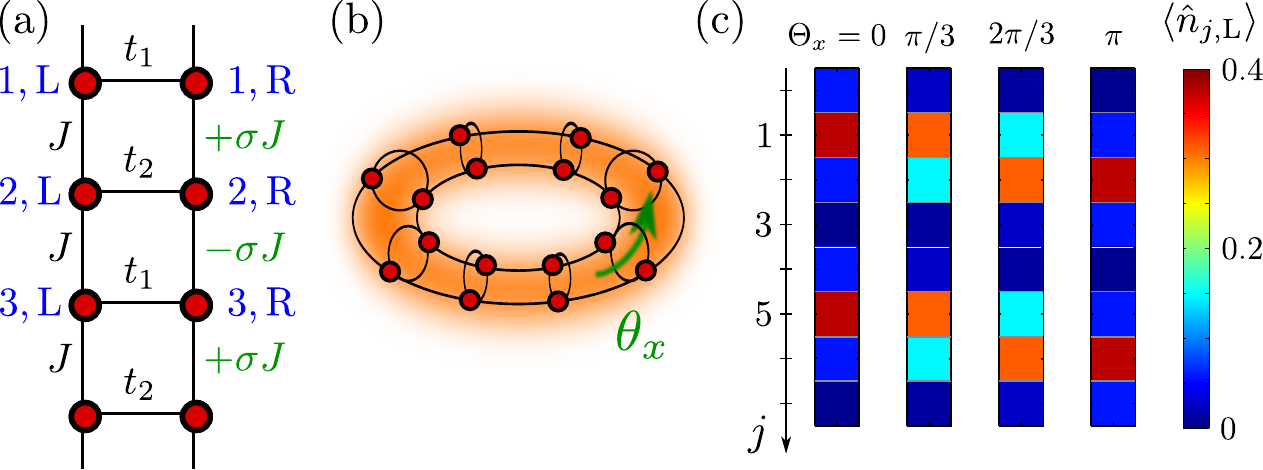, width=\columnwidth}
\caption{(Color online) Using commensurate optical lattice potentials, interacting bosons in 1D ladder systems (a) can be realized. When the hopping elements across the ladder alternate between values $t_1$ and $t_2$, see Eq.\eqref{eq:t1t2Def}, and every second hopping on one leg along the ladder has a phase $\pi$, the thin torus limit of the 2D Hofstadter Hubbard model (at flux per plaquette $\alpha=1/4$) can be realized (b). Which of the hoppings on the right leg have a non-trivial phase $\pi$ depends on the value of $\sigma=\pm1$. The boundary conditions are periodic across the ladder, with a tunable twist angle $\theta_x$, corresponding to magnetic flux $\theta_x/2\pi$ threading the smaller perimeter of the torus. The ground state is an incompressible CDW with average occupation $\rho=1/8$ per lattice site, related to the $1/2$ Laughlin state of the 2D model. The density distribution $\langle\hat{n}_{j,\mathrm{L}}\rangle=\langle\hat{n}_{j,\mathrm{R}}\rangle$ is shown in two unit-cells, chosen from a larger system with open boundary conditions, for different values of $\theta_x$ in (c). This corresponds to a quarter-cycle of the fractionally quantized Thouless pump.}
\label{fig:Intro}
\end{figure}
%%%%%%%%%%%%%%%%%%%%%%%%%%%%%%%%%%%%%%%%%%%%%%%%%%%%%

In this paper we propose a setup for the realization of topological states in strongly interacting bosons, see FIG.\ref{fig:Intro}. To this end, we consider the thin-torus-limit \cite{Bergholtz2005,Bergholtz2006,Bernevig2012} of a 2D fractional Chern insulator  on a square lattice \cite{Sorensen2005,Hafezi2007}, and show how it can be implemented in current experiments with ultracold quantum gases. The resulting groundstate is a CDW at average filling $\rho=1/8$ per lattice site, related to the $\nu=1/2$ Laughlin state \cite{Laughlin1983} in the 2D limit. It can be interpreted as well as a symmetry protected topological phase \cite{Bernevig2012} (protected by inversion symmetry). In addition our model includes the possibility of twisted boundary conditions around the short perimeter of the torus, with a fully tunable twist angle $\theta_x$, see FIG.\ref{fig:Intro} (b). Adiabatically changing this twist angle by $2 \pi$ realizes a many-body version of a Thouless pump \cite{Thouless1983,Berg2011}, which is \emph{fractionally} quantized, see FIG.\ref{fig:Intro} (c).

The paper is organized as follows. In Sec.\ref{sec:Model} we introduce the model and show that it is identical to the thin-torus-limit of the 2D Hofstadter-Hubbard model. Possible experimental realizations are discussed. In Sec.\ref{sec:TopologyNonInteractingSystem} we elucidate on the topological properties in the non-interacting case, and show that they enable a realization of a Thouless pump. In Sec. \ref{sec:InteractingTopology} we return to the discussion of interacting bosons, where DMRG results for the melting of the CDW at weak interactions and the system in a harmonic trap are presented. The topological classification of the ground state is carried out in Sec. \ref{sec:TopClass}, before we close our discussion with a summary and an outlook in Sec. \ref{sec:OutlookSummary}.

%%%%%%%%%%%%%%%%%%%%%%%%%%%%%%%%%%%%%%%%%%%%%%%%%%%%%
%%%%%%%%%%%%%%%%%%%%%%%%%%%%%%%%%%%%%%%%%%%%%%%%%%%%%
\section{Model}
\label{sec:Model}
%%%%%%%%%%%%%%%%%%%%%%%%%%%%%%%%%%%%%%%%%%%%%%%%%%%%%
% sorry for the confusion, I must have changed it looking at Fig 2
%%%%%%%%%%%%%%%%%%%%%%%%%%%%%%%%%%%%%%%%%%%%%%%%%%%%%
We consider the following Bose-Hubbard type model of bosons hopping between the links of a 1D ladder, see FIG.\ref{fig:Intro} (a),
\begin{multline}
\H = - J \sum_{j=1}^L \l \ad_{j+1,\L} \a_{j,\L} + \sigma (-1)^j \ad_{j+1,\R} \a_{j,\R} + \hc \r \\
- \sum_{n=1}^{L/2} \l t_1 \ad_{2n-1,\L} \a_{2n-1,\R} + t_2 \ad_{2n,\L} \a_{2n,\R} + \hc \r \\
+ \frac{U}{2} \sum_{j=1}^L\sum_{\mu=\L,\R} \ad_{j,\mu} \a_{j,\mu} \l  \ad_{j,\mu} \a_{j,\mu}  - 1 \r \\
+V \sum_{j=1}^L \sum_{\mu = \L,\R} \l j - (L+1)/2 \r^2 \ad_{j,\mu} \a_{j,\mu}.
\label{eq:HamiltonianExp}
\end{multline}
Here $\a_{j,\mu}$ annihilates a boson on the left ($\mu=\L$) or the right ($\mu=\R$) leg of the ladder, at the horizontal link $j$. The first line describes hopping along the ladder (vertical links) with amplitude $J$. On the right leg an additional phase $\pi$ is picked up on every second bond, for $\sigma=+1$ from even $j$ to odd $j+1$ and for $\sigma =-1$ from odd $j$ to even $j+1$. Along the horizontal bonds the tunneling rates are alternating between $t_1$ and $t_2$, both assumed to be real-valued and positive. In the third line we added on-site Hubbard-type interactions of strength $U$ everywhere. The model is completed by an external harmonic trapping potential in the fourth line.

Below we will show that the model \eqref{eq:HamiltonianExp} is equivalent to the thin-torus-limit of the 2D Hofstadter Hubbard model. Afterwards we discuss possible experimental realizations with ultra-cold atoms using currently available technology, based on the experimental setups described in \cite{Aidelsburger2011,Aidelsburger2013,Miyake2013,Aidelsburger2014}.

%%%%%%%%%%%%%%%%%%%%%%%%%%%%%%%%%%%%%%%%%%%%%%%%%%%%%
\subsection{Relation to the thin-torus-limit of the Hofstadter-Hubbard model}
\label{subsec:RelationThinTorus}
\begin{figure}[t]
\centering
\epsfig{file=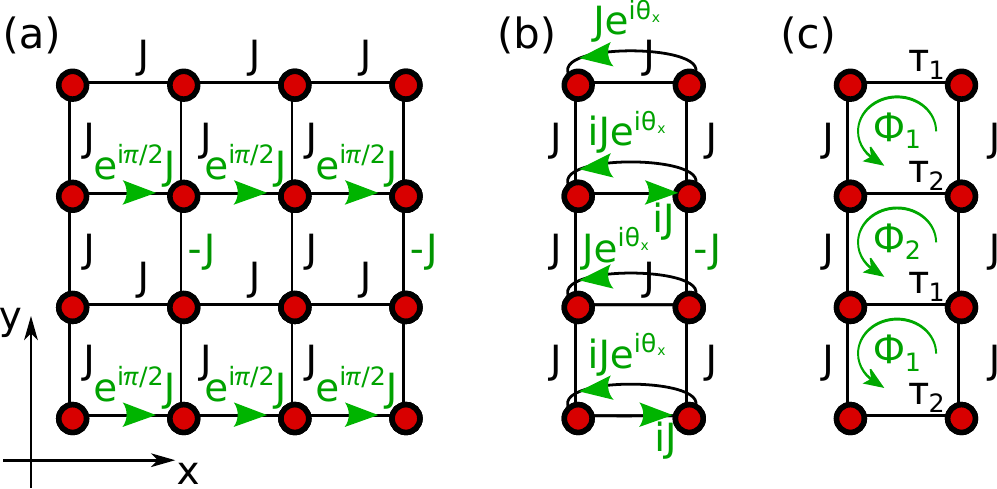, width=0.45\textwidth}
\caption{(Color online) (a) For the 2D Hofstadter Hubbard model at flux per plaquette $\alpha=1/4$ we make a gauge choice leading to a two-by-two magnetic unit-cell. Supplemented by twisted boundary conditions in $x$-direction (twist-angle $\theta_x$), an effective 1D ladder model is obtained when the thin-torus-limit is considered (b). For notational simplicity, the imaginary unit $i = e^{i \pi/2}$ is used to express the complex hopping elements in (b). When an additional gauge transformation is applied, leaving invariant the magnetic flux $\Phi_{1,2}$ in every plaquette, the ladder model described by Eq.\eqref{eq:HamiltonianExp} is obtained (c).}
\label{fig:MappingThinTorus}
\end{figure}

%%%%%%%%%%%%%%%%%%%%%%%%%%%%%%%%%%%%%%%%%%%%%%%%%%%%%
Now we show how the model \eqref{eq:HamiltonianExp} can be related to the thin-torus-limit of the 2D Hofstadter-Hubbard model with flux per plaquette $\alpha=1/4$ (in units of the flux quantum). The latter is described by a Hamiltonian \cite{Sorensen2005,Hafezi2007}
\begin{equation}
\H_{\text{HH}}= - J \sum_{\langle \vec{i}, \vec{j} \rangle } \l \ad_{\vec{i}} \a_{\vec{j}} e^{- i \phi_{\vec{i},\vec{j}}} + \hc \r + \frac{U}{2} \sum_{\vec{j}} \n_{\vec{j}} \l  \n_{\vec{j}}  - 1 \r,
\end{equation}
where $\n_{\vec{j}} = \ad_{\vec{j}} \a_{\vec{j}}$ and $\phi_{\vec{i},\vec{j}}$ are Peierls phases picked up when hopping from site $\vec{j}$ to a neighboring site $\vec{i}$. When a particle is hopping around a plaquette these phases sum up to $\pi/2$, which can be achieved e.g. with the gauge choice $\phi_{\vec{i},\vec{j}}$ shown in FIG.\ref{fig:MappingThinTorus} (a). Next we consider this model  on a torus of size $L_x \times L_y$ with twisted boundary conditions along $x$, i.e. $\psi(x_m+L_x)=e^{i \theta_x} \psi(x_m)$ where $x_m$ is the coordinate of the $m$-th particle, $m=1,...,N$. Such boundary conditions can be implemented by adding additional phases $\theta_x$ to the hoppings from $(L_x,j_y)$ to $(1,j_y)$ for all $j_y=1,...,L_y$. 

%%%%%%%%%%%%%%%%%%%%%%%%%%%%%%%%%%%%%%%%%%%%%%%%%%%%%
%%%%%%%%%%%%%%%%%%%%%%%%%%%%%%%%%%%%%%%%%%%%%%%%%%%%%

We perform the thin-torus-limit by setting the length $L_x=2$ equal to two lattice sites, yielding an effective ladder system as shown in FIG.\ref{fig:MappingThinTorus} (b). Because of the periodic boundary conditions along $x$ there are two possibilities how a boson can tunnel from the left to the right leg of the ladder, originating from paths across the top and bottom of the torus. When summing  up these contributions, we obtain complex hoppings across the ladder $\tau_1=J (1 + e^{-i\theta_x})$ from $(2n-1,\L)$ to $(2n-1,\R)$ and $\tau_2= i J (1 - e^{-i\theta_x})$ from $(2n,\L)$ to $(2n,\R)$ (site-labels as in Eq.\eqref{eq:HamiltonianExp}), see FIG.\ref{fig:MappingThinTorus} (c). The hopping elements along the legs of the ladder are real and given by $-J$ at links $(2n-1,\R)$ to $(2n,\R)$ and $J$ at all other links. 

To show the equivalence of the thin-torus Hofstadter Hubbard model to the Hamiltonian \eqref{eq:HamiltonianExp} we now define $t_{1,2}:=|\tau_{1,2}|$, yielding
\begin{equation}
t_1= J \sqrt{2 \l 1 + \cos \theta_x \r}, \quad  t_2= J \sqrt{2 \l 1 - \cos \theta_x \r},
\label{eq:t1t2Def}
\end{equation}
such that the absolute values of all hopping amplitudes in both models coincide. Thus, we only have to calculate the magnetic fluxes through each of the plaquettes of the ladder and show that they coincide in both models. In the thin-torus-limit of the 2D model we obtain fluxes $\Phi_{1}=(1 + \text{sign} \sin \theta_x)/4$ and $\Phi_{2}=(1 - \text{sign} \sin \theta_x)/4$ in units of the magnetic flux quantum, see FIG.\ref{fig:MappingThinTorus} (b), (c). Choosing 
\begin{equation}
\sigma= -\text{sign} \sin \theta_x
\end{equation}
in Eq.\eqref{eq:HamiltonianExp} we obtain the same fluxes in the 1D ladder model. Therefore the thin-torus-limit of the $\alpha=1/4$ Hofstadter Hubbard model is equivalent to the model \eqref{eq:HamiltonianExp}, up to a unitary gauge transformation $\hat{U}(\theta_x)$ which depends explicitly on the twist-angle $\theta_x$.

Quasi 1D ladder systems were investigated before, and it was shown that effective gauge fields give rise to interesting physical effects such as Meissner currents \cite{Petrescu2013,Atala2014}. A version of the thin-torus-limit of the Hofstadter model at arbitrary flux per plaquette $\alpha$ was also studied already in \cite{Hugel2013}. In contrast to our model, this work did not take into account periodic boundary conditions across the ladder, resulting in a \emph{homogeneous} flux $\Phi_{1} = \Phi_2 = \alpha$ per plaquette and equal hopping amplitudes corresponding to $|t_1|=|t_2|=|J|$ using our notations. As a consequence the authors could study interesting edge states, carrying chiral currents along opposite edges of the ladder. Within our model on the other hand, we can study the effect of tunable twisted boundary conditions across the ladder. In addition, we study interactions between bosons.

%%%%%%%%%%%%%%%%%%%%%%%%%%%%%%%%%%%%%%%%%%%%%%%%%%%%%
\subsection{Possible experimental implementation}
%%%%%%%%%%%%%%%%%%%%%%%%%%%%%%%%%%%%%%%%%%%%%%%%%%%%%
Next we discuss a possible experimental realization of our scheme. The first required ingredient is a superlattice for creating ladders (with a four-site unit-cell), which has been implemented experimentally, see e.g. \cite{Foelling2007,Atala2014}. The second ingredient is a (staggered) artificial gauge field, which has been experimentally demonstrated as well \cite{Aidelsburger2011,Aidelsburger2013,Miyake2013,Jotzu2014,Aidelsburger2014}. The implementation of the Hamiltonian \eqref{eq:HamiltonianExp} is motivated and very closely related to the recent experiment \cite{Aidelsburger2014}, and we think that alternative realizations should be possible.

%%%%%%%%%%%%%%%%%%%%%%%%%%%%%%%%%%%%%%%%%%%%%%%%%%%%%
\begin{figure}[t]
\centering
\epsfig{file=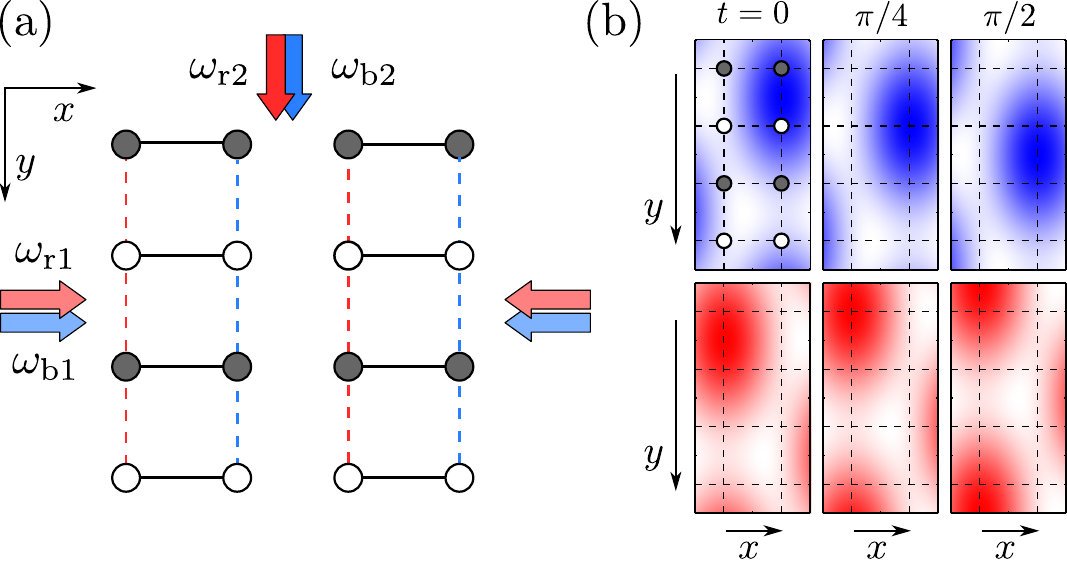, width=0.9\columnwidth}
\caption{(Color online)(a) Possible realization of the ladder system with half a magnetic flux-quantum piercing every second plaquette, cf. \cite{Aidelsburger2014}. By interference of  standing waves from a red ($r$) and a blue ($b$) detuned sideband of the long-lattice laser in $x$ direction, with corresponding slightly detuned running waves along $y$ direction, two independent lattice modulations (upper blue and lower red plot in (b)) are created. Each acts on a single leg of the ladder and they move in opposite directions along $y$,  as shown by the amplitude of the modulations for different times in (b).}
\label{fig:realization}
\end{figure}
%%%%%%%%%%%%%%%%%%%%%%%%%%%%%%%%%%%%%%%%%%%%%%%%%%%%%

To begin with, our scheme requires a cubic lattice created by standing waves with short wavelength $\lambda_S$ both in $x$ and $y$ direction. We choose the origin such, that lattice sites are centered at $x_{j,\mathrm{L}}=0$, $x_{j,\mathrm{R}}=\lambda_S/2$, $y_{j, \mu}=(j-1)\lambda_S/2$.  Additional standing waves with long wavelength $\lambda_L=2\lambda_S$ are required in both directions. The strong long-lattice along $x$ separates the individual ladders, whereas the weaker long-lattice along $y$ induces a staggered potential of strength $\Delta$ along the legs that is indicated by white and grey filled sites in Fig. \ref{fig:realization}(a). This staggered potential is required for realizing the artificial magnetic field. We will denote the bare hopping elements in the so-obtained ladder by $J_y$ (along the ladder) and $J_x$ (across the ladder).

The alternating flux $\Phi_{1,2}=0,1/2$ (in units of the magnetic flux quantum) can be realized with a similar configuration as in the experiment \cite{Aidelsburger2014}, where a homogeneous flux of $\alpha=1/4$ was realized via laser assisted tunneling \cite{Jaksch2003}. Bare hopping along the legs is strongly suppressed by the staggered potential $\Delta \gg J_y$, and has to be restored by resonant modulation of the potential landscape. This can be achieved by a time-dependent potential of the form $V(x,y,t) = V_0 \cos (\Delta t + g_{x,y})$, and as pointed out by Kolovsky \cite{Kolovsky2011a} the freedom in choosing the phase-shifts $g_{x,y}$ allows to implement Peierls phases -- and thus to create artificial gauge fields. To implement a suitable time-dependent potential experimentally, two side-bands of the long-wavelength laser can be employed. They make up four additional beams, two red-detuned ones with frequencies $\omega_{r1}$ and $\omega_{r2}$, and two blue-detuned ones at frequencies $\omega_{b1}$ and $\omega_{b2}$. When the red sidebands are sufficiently far detuned from the blue sidebands, i.e. $\omega_{rj} - \omega_{bj} \gg \Delta$ for both $j=1,2$, interference terms between them can be neglected and they can be treated separately form each other. 

We now move on by constructing suitable interference patterns between the red-detuned and blue-detuned pairs of beams respectively, with relative frequencies $\omega_i = \omega_{i2}-\omega_{i1}$ where $i=r,b$. These beat-notes give rise to the required modulation of the potential at frequency $\Delta$, and we chose them to be $\omega_r = -\Delta$, $\omega_b= \Delta$.
As in \cite{Aidelsburger2014} both beams $r1$ and $b1$ are retroreflected in $x$ direction to form standing waves, see FIG. \ref{fig:realization} (a), and they interfere with running waves $r2$ and $b2$ in $y$ direction. This configuration gives rise to the time-dependent interference patterns shown in FIG.\ref{fig:realization} (b)
\begin{align}
V_r(x,y,t) &= V_0/4\big[ 1+4 \cos^2(k_L x)\nonumber\\
&+4 \cos(k_L x)\cos(k_L y+\Delta t-\pi/4) \big],\\
V_b(x,y,t) &= V_0/4\big[ 1+4 \sin^2(k_L x)\nonumber\\
&+4 \sin(k_L x)\cos(k_L y-\Delta t-\pi/4) \big].
\label{eq:Vxyt}
\end{align}
From FIG.\ref{fig:realization} (b) we recognize that the phase of the retroreflected red sideband is chosen such that the resulting modulation is restricted to the left leg of the ladder and moves in negative $y$ direction in time. The blue sideband, vice-versa, leads to a modulation restricted to the right leg of the ladder which is moving in positive $y$-direction in time. This counter-directed movement of the potential modulation introduces angular momentum into the system, which mimics the effect of a magnetic field. Note that the additional standing waves in Eq.\eqref{eq:Vxyt} sum up to a constant overall energy shift only. The described setup is completely analogous to the one implemented in \cite{Aidelsburger2014}, except for the phases chosen for the different laser beams.

Now, as desired, every lattice site is subject to a time-dependent modulation of the local potential $V_{j, \mu}=V_0\cos(\Delta t + g_{j,\mu})$ (with $\mu = \text{L}, \text{R}$). From Eq.\eqref{eq:Vxyt} we read off the phase shifts, which are given by $g_{j,L}=-3\pi/4+j\pi/2$ and $g_{j,R}=3\pi/4-j\pi/2$. To proceed and calculate the resulting Peierls phases, let us consider the simplified case when two lattice sites $1$ and $2$ are coupled by a hopping element $J_y$, where the second site is detuned by an energy $\Delta \gg J_y$ from the first one. Resonant periodic modulations of the local potentials $V_1, V_2$ with frequency $\Delta$ and phases $g_1$ and $g_2$ restore strong hopping. Indeed, the effective tunneling matrix element from $1$ to $2$ is given by $J_\text{eff}=J(e^{-i g_1}-e^{-i g_2}) / \sqrt{2}$ \cite{Aidelsburger2014,Kolovsky2011a}, where we defined the amplitude $J$ as
\begin{equation}
J = J_y V_0/(\sqrt{2}\Delta).
\end{equation}
Returning to the ladder model, in this way we find for the induced hoppings
\begin{align}
J_{(j,L),(j+1,L)}&=-e^{ij\pi}J_y V_0/(2\Delta)(e^{i g_{j,L}}-e^{i g_{j+1,L})}\nonumber\\
&=J e^{ij\pi/2},\\
J_{(j,R),(j+1,R)}&=-e^{ij\pi}J_y V_0/(2\Delta)(e^{i g_{j,L}}-e^{i g_{j+1,L})}\nonumber\\
&=J e^{-ij\pi/2}.
\end{align}
The above configuration can be mapped to Eq.\eqref{eq:HamiltonianExp} via a gauge transformation. 

Finally we turn to the implementation of the hoppings $t_{1,2}$ connecting the legs of the ladder. Without further modifications of the described setup, they are given by $t_1=t_2=J_x$. Choosing the modulation strength $V_0$ such that $J=J_x/\sqrt{2}$ readily realizes the cases $\theta_x=\pm  \pi/2$. They are of special relevance, because the resulting model is inversion symmetric around the center of links on the legs. As will be shown below, the model supports inversion-symmetry protected topological phases at these points. In order to realize arbitrary values of $\theta_x$, the hoppings $t_{1,2}$ can be manipulated with a second independent square lattice rotated by $45^\circ$, such that the potential barrier along every second horizontal bond of the ladder is increased when the lattice is properly adjusted. To this end an additional sideband of the short wavelength laser is added in $x$ and $y$ direction and both beams are retroreflected. The resulting interference pattern realizes the required rotated square lattice with lattice constant $\lambda_{S}/\sqrt{2}$. Lastly we note that the periodic modulation used to restore hoppings along the ladder reduces the tunneling amplitudes $t_{1,2}$ between the legs. This gives rise to effective hoppings $t_{1,2}^\mathrm{eff}/t_{1,2}=1-(2-\sqrt{2}) V_0^2/ (4 \Delta^2)$, but the effect can be neglected for large $\Delta$.

%%%%%%%%%%%%%%%%%%%%%%%%%%%%%%%%%%%%%%%%%%%%%%%%%%%%%
\section{Topology in the non-interacting system -- Thouless pump}
\label{sec:TopologyNonInteractingSystem}
%%%%%%%%%%%%%%%%%%%%%%%%%%%%%%%%%%%%%%%%%%%%%%%%%%%%%
We start the analysis of our model by investigating non-interacting bosons. In this case all properties of the bandstructure immediately follow from the 2D Hofstadter model \cite{HOFSTADTER1976}. The lowest band of the 2D Hofstadter Hamiltonian at $\alpha=1/4$ is characterized by a Chern number $C=1$, which gives rise to a quantized Hall current perpendicular to an applied external force. We show below that such quantized particle transport along the 1D ladder survives in the thin-torus-limit, when the external force is induced by inserting magnetic flux through the smaller perimeter of the torus. Experimentally this corresponds to an adiabatic change of the twisted boundary conditions, $\partial_t \theta_x \neq 0$. A change of $\theta_x$ by $2 \pi$ can also be interpreted as one cycle of a Thouless pump \cite{Thouless1983}.

Now we discuss the relation between Bloch wavefunctions of the one- and two-dimensional models. At $\alpha=1/4$ the 2D Hofstadter model has a four-site unit-cell. Using the gauge choice introduced earlier in FIG.\ref{fig:MappingThinTorus} (a), we can calculate the Bloch Hamiltonian $\H(k_x,k_y)$ of the 2D model, resulting in Bloch wavefunctions $\ket{u(k_x,k_y)}$ (see e.g. Supplementary Material in \cite{Aidelsburger2014} for a concrete calculation). When performing the thin-torus-limit as described in \ref{subsec:RelationThinTorus}, the Bloch wavefunctions do not change, except that the quasimomentum $k_x$ across the resulting ladder is replaced by the angle $\theta_x$ defining the twisted periodic boundary conditions. The Bloch wavefunction in the thin-torus-limit thus reads $\ket{u(\theta_x,k_y)}$. By applying the gauge transformation $\hat{U}(\theta_x)$ (see \ref{subsec:RelationThinTorus}) we also obtain the Bloch function of the 1D ladder model \eqref{eq:HamiltonianExp}, $\ket{v(\theta_x;k_y)} = \hat{U}(\theta_x) \ket{u(\theta_x,k_y)}$, at a given twist angle $\theta_x$. Consequently the Bloch bands $\epsilon_n(k_x,k_y)$ labeled by $n=1,...,4$ (i.e. the eigenenergies of the Bloch-Hamiltonian) of the 2D Hofstadter Hubbard model coincide with those of the 1D ladder model \eqref{eq:HamiltonianExp}, $\epsilon_n(\theta_x;k_y)$.

From the Bloch wavefunctions we will now derive the topological properties of the 1D ladder model \eqref{eq:HamiltonianExp}. To this end we calculate the Zak phase \cite{Zak1989} $\varphi_{\text{Zak}}(\theta_x)$ for a path through the Brillouin zone (BZ) along $k_y$ and for a given value of $\theta_x$. Because the Zak phase is invariant under the gauge-transformation $\hat{U}(\theta_x)$, the 1D ladder model reproduces the result $\varphi_{\text{Zak}}(\theta_x)|_{\text{H}}$ of the 2D Hofstadter model, $\varphi_{\text{Zak}}(\theta_x) = \varphi_{\text{Zak}}(\theta_x) |_{\text{H}}$. Zak phases can be measured in ultra cold atom systems using Ramsey interferometry in combination with Bloch oscillations \cite{Atala2012}. A similar measurement in the simplified 1D model would thus allow to study the Berry curvature of the 2D Hofstadter model, see also \cite{Abanin2012,Duca2014}.

A characteristic feature of the Hofstadter model at $\alpha=1/4$ is that its lowest band is topologically non-trivial, with a Chern number $C_\text{H}=1$. The Chern number can be directly related to the winding of the Zak phase \cite{Xiao2010},
\begin{equation}
C_{\text{H}} = \frac{1}{2 \pi} \int_{\text{BZ}} dk_x  ~ \partial_{k_x} \varphi_{\text{Zak}}(k_x) |_\text{H}.
\end{equation}
For the 1D ladder model the winding of the Zak phase equivalently defines the Chern number,
\begin{equation}
C =  \frac{1}{2 \pi} \int_{0}^{2\pi} d\theta_x  ~ \partial_{\theta_x} \varphi_{\text{Zak}}(\theta_x).
\label{eq:defChernTP}
\end{equation}
Now we discuss the physical consequences of the non-trivial Chern numbers $C=C_\text{H} =1 $. In the case of the 2D Hofstadter model, it is related to the Hall current induced by a constant external force \cite{Thouless1982,Vonklitzing1980}. This current was recently measured with essentially non-interacting atoms, in ultra cold Fermi gases \cite{Jotzu2014} and also with ultra cold bosons homogeneously populating the lowest Bloch band \cite{Aidelsburger2014}. In the thin-torus limit of the Hofstadter model, a constant force around the short perimeter of the torus can be applied by adiabatically changing the twist angle in the boundary conditions, $F \propto \partial_t \theta_x$. Like in the 2D model, this leads to a Hall current perpendicular to the induced force -- i.e. along the ladder -- which is quantized and proportional to the Chern number $C$. 

An alternative interpretation of the quantized current in the 1D model \eqref{eq:HamiltonianExp} is given by the concept of a Thouless pump \cite{Thouless1983}. To understand this, we note that the Zak phase is related to the macroscopic polarization $P = a \varphi_{\text{Zak}} / 2 \pi$, where $a$ is the extent of the magnetic unit-cell in $y$-direction \cite{Kingsmith1993}. Now, by definition \eqref{eq:defChernTP}, it follows that the Zak-phase changes continuously from $0$ to $C \times 2 \pi$ when the parameter $\theta_x$ is adiabatically changed by $2 \pi$ in a time $T$. This corresponds to a quantized change of the polarization by $\Delta P = C a$, or a quantized current $C a / T$. Below (in \ref{sec:TopClass}) we give an intuitive explanation of the microscopic mechanism of this effect in the 1D ladder system.

Experimentally the Thouless pump could be detected by loading non-interacting ultra cold atoms (bosons or fermions) into the lowest Bloch band. Then, comparing in-situ images of the atomic cloud before and after adiabatically changing $\theta_x$ by $2 \pi$ reveals the quantized current. This is similar to the measurements performed recently on the 2D Hofstadter model \cite{Jotzu2014,Aidelsburger2014}. Now we turn to the discussion of interacting atoms, where we will give an example for a \emph{fractionally quantized} Thouless pump corresponding to a Chern number $C=1/2$.

%%%%%%%%%%%%%%%%%%%%%%%%%%%%%%%%%%%%%%%%%%%%%%%%%%%%%
\section{Interacting topological states}
\label{sec:InteractingTopology}
%%%%%%%%%%%%%%%%%%%%%%%%%%%%%%%%%%%%%%%%%%%%%%%%%%%%%

As a next step we add local Hubbard-type interactions between the bosons to the investigation of our model. In the 2D limit of the Hofstadter-Hubbard model the existence of an incompressible Laughlin-type ground state has been established numerically for fluxes within a range $\alpha=0...0.4$ \cite{Sorensen2005,Hafezi2007}. For $\alpha=1/4$ studied in this paper we thus expect a fractional Chern insulator at a magnetic filling $\nu=N/N_\phi=1/2$, where $N$ is the number of particles and $N_\phi$ the number of flux quanta in the system. The average occupation number of each lattice site is thus $\rho=1/8$ in this phase. Now we will show using DMRG calculations that in the thin-torus limit an incompressible CDW survives at the same filling. We study the robustness of this phase when the interaction strength $U$ is lowered. In a harmonic trap the incompressible phase is shown to be robust enough to form plateaus of constant density.

%%%%%%%%%%%%%%%%%%%%%%%%%%%%%%%%%%%%%%%%%%%%%%%%%%%%%
\subsection{Grand-canonical phase diagram}
%%%%%%%%%%%%%%%%%%%%%%%%%%%%%%%%%%%%%%%%%%%%%%%%%%%%%

We use a Matrix Product State (MPS) based algorithm to find the ground state of finite size ladder systems with open boundary conditions (obc) \cite{Schollwoeck2011}. MPS are very well suited to approximate the CDW like states we expect in the incompressible phase and can -- with increased resources -- also describe the melting of the CDW at fillings near $\rho=1/8$. By varying the chemical potential we have determined the ground state energy for particle numbers around $N \approx L_y/4$ (corresponding to $\rho=1/8$) and three different interaction energies $U/J=2, 5, \infty$. Due to symmetries it is sufficient to consider twist angles from the parameter space $\theta_x\in [0,\pi/2]$. 

\begin{figure}[t]
\centering
\epsfig{file=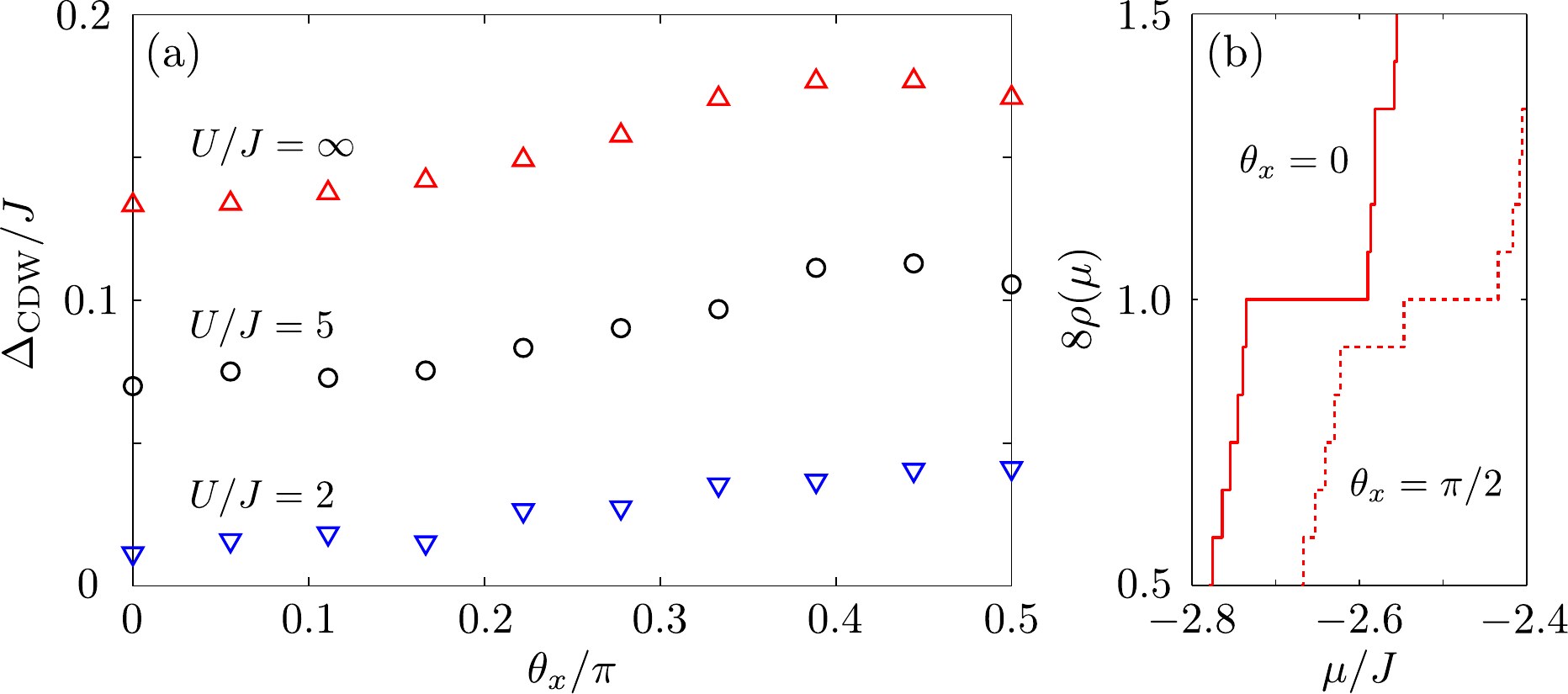, width=\columnwidth}
\caption{(Color online) (a) The particle-hole gap $\Delta_\text{CDW}$ of the incompressible phase at filling $\rho=1/8$ for varying interaction strength $U/J=\infty, 5,2$ extrapolated from finite system size calculations at $L_y=18,24,32$. This gap corresponds to the plateaus in the $\rho(\mu)$ diagrams shown in (b) for $U/J=\infty$ at system size $L_y=48$. Note that for $\theta_x=\pi/2$ the plateau has a kink in its middle where $\rho(\mu)$ changes, corresponding to the addition of a single particle. This is not a bulk effect, however, because the additional particle is localized at the edge of the system.}
\label{fig:GrandCanonicalPD}
\end{figure}

We now define the critical chemical potentials $\mu_{1/8\pm}$ as the upper and lower boundaries of the incompressible CDW phase. In Fig. \ref{fig:GrandCanonicalPD}(a) we show the corresponding particle hole gap $\Delta_\text{CDW}=\mu_{1/8+}-\mu_{1/8-}$  extrapolated to thermodynamic limit from finite system results at $L_y=16,24,32$. Strong interactions stabilize the non trivial CDW that is protected by a gap on the order of $\Delta_\text{CDW}^{U=\infty}\approx J/6$. At moderate interaction $U=5$ the incompressible phase is still protected by $\Delta_\text{CDW}^{U=5}\approx J/12$, whereas for $U=2$ the gap almost closes and the CDW phase vanishes. 

The topological nature of our system and the presence of obc edges have to be taken into account when analysing dependence of the particle number $N(\mu)$ on the chemical potential, as shown in Fig. \ref{fig:GrandCanonicalPD}(b). At $\theta_x=0$ we find a single plateau at filling $\rho=1/8$, however at $\theta_x=\pi/2$ this plateu is split by the addition of a single particle at intermediate chemical potential. This is an edge effect and strongly dependent on the choice of boundary conditions \cite{Grusdt2013EdgeStates}, allowing us to interpret the full plateau as an incompressible bulk phase.

%%%%%%%%%%%%%%%%%%%%%%%%%%%%%%%%%%%%%%%%%%%%%%%%%%%%%
%%%%%%%%%%%%%%%%%%%%%%%%%%%%%%%%%%%%%%%%%%%%%%%%%%%%%

%%%%%%%%%%%%%%%%%%%%%%%%%%%%%%%%%%%%%%%%%%%%%%%%%%%%%
\subsection{Harmonic trapping potential}
%%%%%%%%%%%%%%%%%%%%%%%%%%%%%%%%%%%%%%%%%%%%%%%%%%%%%

The incompressible, integer filling phases of the conventional Bose-Hubbard model can be nicely demonstrated in harmonically trapped systems where Mott-insulating plateaus of constant density emerge, surrounded by superfluid regions ("wedding-cake" structure). We here show that in a similar fashion the non trivial CDW phase on the ladder could be visualized in harmonically trapped gases. 

\begin{figure}[t]
\centering
\epsfig{file=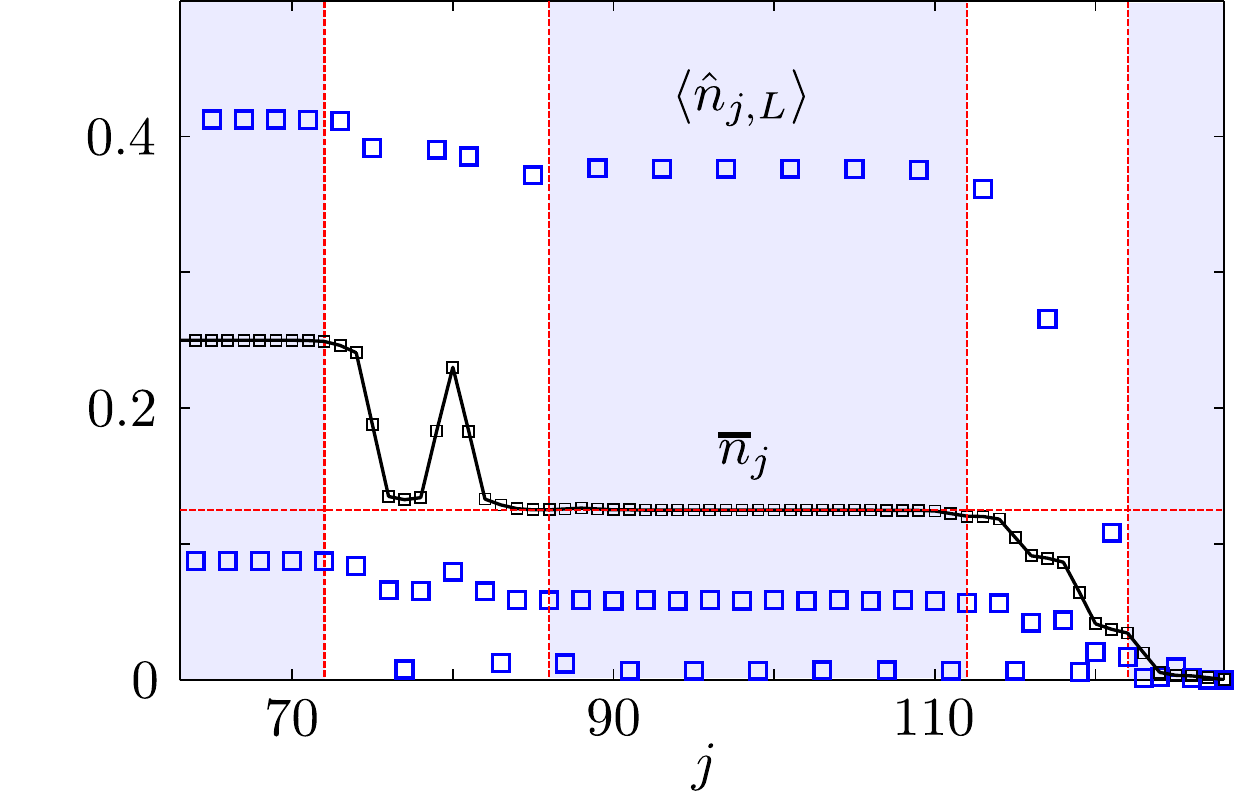, width=0.80\columnwidth} $\qquad$
\caption{(Color online) Local density of hard-core bosons with $U/J=\infty$ at $\theta_x=0$ in a harmonic trap centered around $j=64.5$ with $V=8.4\times 10^{-5} J$ and $\mu=-2.4J$ and $L_y=128$. While the density along the left leg (blue squares) demonstrates the density wave character, the averaged density (black solid line) reveals incompressible phases at fillings $\rho=1/4$ and $\rho=1/8$. The vertical red lines indicate the phase boundaries between compressible and incompressible (blue shading) phases in local density appriximation}
\label{fig:harmonicTrap}
\end{figure}

Using the MPS code (with increased bond dimension to correctly describe the compressible regions in the trap) we have calculated the density distribution in traps up to size $L_y=128$ for fixed global chemical potential $\mu$. The trap depth $V$ is choosen such that from local density approximation we expect a wedding cake structure of quarter filling in the center, a compressible transition region and a large incompressible region of filling $\rho=1/8$ before vacuum. As we show in Fig. \ref{fig:harmonicTrap} this picture is well reproduced by the numeric simulation, where the local density $\langle \hat{n}_{j,L} \rangle$ reveals the CDW nature. To check the incompressibility of the phases, we calculated an averaged density $ \overline{n}_j=1/8\sum_{i=j-1}^{j+2} \langle\hat{n}_{i,\text{L}}+\hat{n}_{i,\text{R}}\rangle$. It illustrates the two density plateaus which lie within the extend predicted by the local chemical potential $\mu(x) = \mu + V_\text{trap}(x)$ and the critical chemical potantials $\mu_{1/8,\pm}$ calculated in the previous section. 

The outer incompressible phase is the CDW state at filling $\rho=1/8$ we are mostly interested in, corresponding to a half-filled lowest Bloch band. The inner incompressible phase at quarter-filling corresponds to a completely filled lowest Bloch band, and is similar to the Mott phase of bosons in the lowest band of the 1D Su-Schrieffer-Heeger model \cite{Grusdt2013EdgeStates}.

%%%%%%%%%%%%%%%%%%%%%%%%%%%%%%%%%%%%%%%%%%%%%%%%%%%%%
%%%%%%%%%%%%%%%%%%%%%%%%%%%%%%%%%%%%%%%%%%%%%%%%%%%%%

%%%%%%%%%%%%%%%%%%%%%%%%%%%%%%%%%%%%%%%%%%%%%%%%%%%%%
%%%%%%%%%%%%%%%%%%%%%%%%%%%%%%%%%%%%%%%%%%%%%%%%%%%%%
\section{Topological classification and fractional Thouless pump}
\label{sec:TopClass}
%%%%%%%%%%%%%%%%%%%%%%%%%%%%%%%%%%%%%%%%%%%%%%%%%%%%%
%%%%%%%%%%%%%%%%%%%%%%%%%%%%%%%%%%%%%%%%%%%%%%%%%%%%%

Now we discuss the topological properties of the $\rho=1/8$ CDW phase. We distinguish two cases for the classification of the phase, the $1+1$D model where the second dimension is defined by the twist-angle $\theta_x=0...2 \pi$, and the $1$D model at points of highest symmetry $\theta_x = \pm \pi/2$. In the first case, robust topological properties carry over from the $\nu=1/2$ LN state from the 2D Hofstadter-Hubbard model. In the second case, the CDW constitutes a (inversion-) symmetry-protected topological phase (SPT) which is not robust against disorder.

%%%%%%%%%%%%%%%%%%%%%%%%%%%%%%%%%%%%%%%%%%%%%%%%%%%%%
\subsection{$1+1$D model and fractional Thouless pump}
%%%%%%%%%%%%%%%%%%%%%%%%%%%%%%%%%%%%%%%%%%%%%%%%%%%%%
The $1/2$ LN state in the 2D Hofstadter-Hubbard model is characterized by a fractionally quantized Chern number $C=1/2$ \cite{Niu1985,Hafezi2007}, and as will be shown shortly this carries over to the $1+1$D gapped CDW state. Before going through the details of the calculation, however, let us give an intuitive physical picture. 

As mentioned above, the Chern number is directly related to the quantized Hall current in the 2D model (on a torus). If one unit of magnetic flux is introduced through the perimeter of the torus, i.e. $\Delta \theta_x=2 \pi$, a quantized Hall current around the torus is induced. While in the case of an integer-quantized Chern number $C=p$ the state returns to itself immediately, when $C=p/q$ takes a fractional value the state returns to itself only after introduction of $q$ flux quanta, $\Delta\theta_x = q \times 2 \pi$. 

\begin{figure}[t]
\centering
\epsfig{file=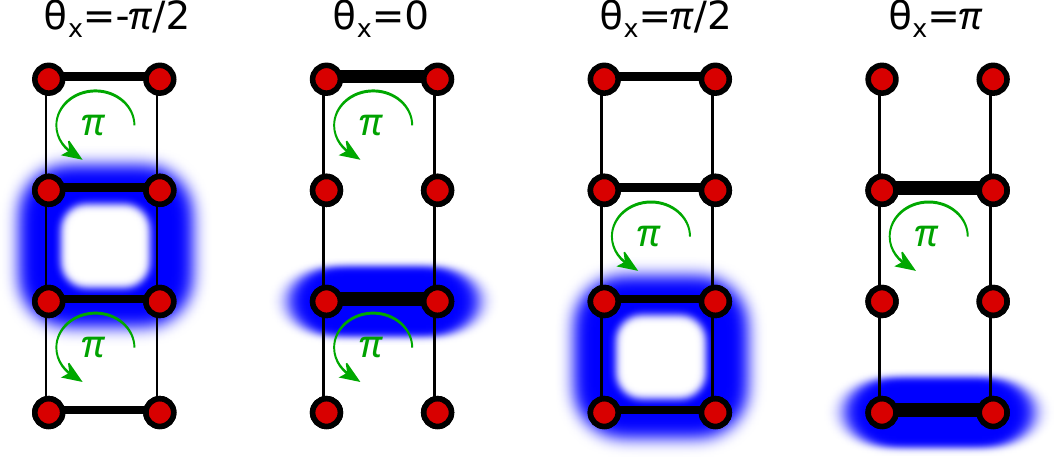, width=0.45\textwidth}
\caption{(Color online) Approximate Wannier orbitals (blue shaded) at the points $\theta_x$ of highest symmetry. }
\label{fig:ThoulessPumpOrbitals}
\end{figure}

As discussed in the non-interacting case, an integer-quantized Thouless pump still exists in the thin-torus limit when the twist-angle $\theta_x$ is adiabatically increased. This mechanism carries over to the $\rho=1/8$ CDW, as can be understood from a simple Gutzwiller-ansatz. To this end we approximate the CDW by a product state
\begin{equation}
\ket{\text{CDW}} = \prod_{n} \bd_{2n}(\theta_x) \ket{0},
\label{eq:CDWgutzwiller}
\end{equation}
where $\bd_j(\theta_x)$ creates a boson in the Wannier orbital corresponding to unit-cell $j$. To understand how the Wannier orbitals depend on $\theta_x$, we approximate them at the points of highest symmetry, $\theta_x = 0, \pi/2 ,\pi ,...$. To this end we search for the state of lowest energy within each unit-cell, and note that in principle the residual coupling between unit-cells could be treated perturbatively. The result is illustrated in FIG.\ref{fig:ThoulessPumpOrbitals}. At $\theta_x=0$ the hoppings are $J$, $t_1=2 J$ and $t_2=0$, such that Wannier orbitals are localized on every other rung, with an energy of $- t_2 = -2 J$ to zeroth order in the described perturbation theory. At $\theta_x=\pm \pi/2$ on the other hand, the hoppings read $J$ and $t_1=t_2=\sqrt{2} J$ such that considering only rungs is not sufficient. Instead we compare the energy of a particle hopping around a single four-site plaquette with zero and $\pi$ flux respectively. While in the latter case there are two degenerate states with energy $- \sqrt{3} J$, for vanishing flux we find a non-degenerate state with lower energy $-\l 1 + \sqrt{2} \r J$.

%%%%%%%%%%%%%%%%%%%%%%%%%%%%%%%%%%%%%%%%%%%%%%%%%%%%%
%%%%%%%%%%%%%%%%%%%%%%%%%%%%%%%%%%%%%%%%%%%%%%%%%%%%%

Although we consider only local Hubbard-type interactions, the CDW state \eqref{eq:CDWgutzwiller} is stabilized by a finite gap $\Delta_\text{CDW}$ to any excitations (it can also be interpreted as a Mott insulator). This is due to a hopping-induced finite range interaction. If we calculate the Wannier orbitals beyond the zeroth order approximation introduced above, nearest and next-nearest neighbor orbitals acquire a finite overlap. Thus, if the twist-angle $\theta_x$ is adiabatically changed, the state \eqref{eq:CDWgutzwiller} follows the modified Wannier orbitals. Because they re-connect to their neighbors after a full pumping cycle, see FIG.\ref{fig:ThoulessPumpOrbitals}, a quantized atomic current flows along the ladder. Because -- assuming periodic boundary conditions along $y$ -- the state only returns to itself after \emph{two} full pumping cycles, the Thouless pump is fractionally quantized, with a coefficient (the Chern number) $C=1/2$. This quantization is robust against any perturbations which are small compared to the gap $\Delta_\text{CDW}$. The Thouless pump is also illustrated in FIG.\ref{fig:Intro} (c).

Now we turn to a more formal topological classification of the $1+1$D model, following \cite{Berg2011}. To this end we calculate the many-body Chern number of the $\rho=1/8$ CDW state. Because periodic boundary conditions are required along the ladder (in $y$-direction), we restrict our analysis in this section to exact diagonalization of small systems (instead of performing DMRG calculations as before). Before starting, we note that on a torus the CDW ground state is two-fold degenerate (in thermodynamic limit $L_y \rightarrow \infty$), as expected from the topologically protected two-fold ground state degeneracy of the $1/2$ LN state in the 2D Hofstadter model. Naively this degeneracy can be understood in 1D from the obvious ambiguity in the choice of occupied orbitals in Eq.\eqref{eq:CDWgutzwiller}: Choosing odd orbitals instead, $\ket{\text{CDW}'}=\prod_n \bd_{2n+1}(\theta_x) \ket{0}$, yields an equivalent but orthogonal CDW state. We saw already in the discussion above, that $\ket{\text{CDW}}$ can be adiabatically transformed into $\ket{\text{CDW}'}$ without closing the bulk gap by applying one full Thouless pumping cycle. 

\begin{figure}[t]
\centering
\epsfig{file=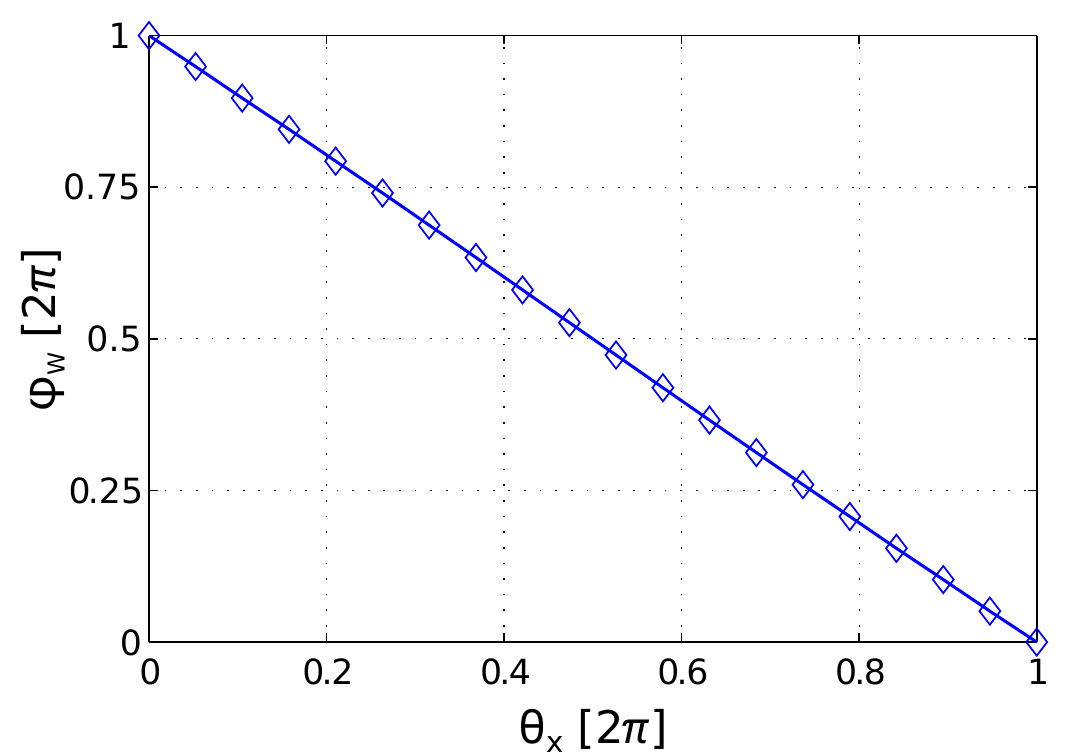, width=0.42\textwidth}
\caption{(Color online) The $U(2)$ Wilson-loop phase $\varphi_\text{W}(\theta_x)= \text{Im} \log \text{det} \hat{W}(\theta_x)$ is shown for the $\rho=1/8$ CDW, the winding of which gives the total Chern number. We used exact diagonalization for a system of size $L_x=2$, $L_y=12$ with periodic boundary conditions and $N=3$ particles for $N_\phi=6$ flux quanta.}
\label{fig:ManyBodyChern}
\end{figure}

The many-body Chern number is defined, in analogy to the single-particle case, by employing twisted boundary conditions. In the $1$D thin-torus limit model we thus have a $2$D parameter space spanned by the external parameter $\theta_x=0...2 \pi$ and the the twist angle $\theta_y=0...2 \pi$ of the $1$D ladder model. However, because the groundstate is two-fold degenerate (for some values $\theta_{x,y}$ this is true even in a finite system), only the \emph{total} Chern number of both states can be defined. It can most conveniently be calculated as the winding of the $U(2)$ Wilson loop $\hat{W}$, which is a non-Abelian generalization of the Zak phase \cite{Wilczek1984,Yu2011}. It is defined via
\begin{equation}
\hat{W}(\theta_x) = \hat{\mathcal{P}} \exp \l - i \int_0^{2 \pi} d \theta_y ~ \hat{\mathcal{A}}(\vec{\theta}) \r,
\end{equation}
where $\hat{\mathcal{A}}(\vec{\theta})$ is the non-Abelian Berry connection \cite{Wilczek1984} and $\hat{\mathcal{P}}$ denotes path-ordering. Then its winding yields the \emph{total} Chern number $C_\text{tot}$, which divided by the number of degenerate states $N_\text{deg}$ -- in our case $N_\text{deg}=2$ -- yields the fractional Chern number,
\begin{equation} 
C = \frac{1}{N_\text{deg}} \times \frac{1}{2 \pi} \int_0^{2 \pi} d \theta_x ~ \partial_{\theta_x} \underbrace{\text{Im} \log \text{det} \hat{W}(\theta_x)}_{=\varphi_W}.
\end{equation}
The Wilson loop phase can easily be evaluated numerically in a gauge-independent way, see e.g. \cite{Yu2011}, and in FIG.\ref{fig:ManyBodyChern} we show $\varphi_W(\theta_x)$ for the thin-torus model. We observe a winding by $2 \pi$, which as expected results in a many-body Chern number $C=2 \pi / (2 \times 2 \pi) = 1/2$.

%%%%%%%%%%%%%%%%%%%%%%%%%%%%%%%%%%%%%%%%%%%%%%%%%%%%%
%%%%%%%%%%%%%%%%%%%%%%%%%%%%%%%%%%%%%%%%%%%%%%%%%%%%%

%%%%%%%%%%%%%%%%%%%%%%%%%%%%%%%%%%%%%%%%%%%%%%%%%%%%%
\subsection{$1$D model and SPT CDW}
%%%%%%%%%%%%%%%%%%%%%%%%%%%%%%%%%%%%%%%%%%%%%%%%%%%%%
At special values of the twist angle $\theta_x=\pm \pi/2$ the model \eqref{eq:HamiltonianExp} is inversion-symmetric around the center of links on the legs of the ladder. In this case, the CDW phase can be understood as a SPT phase \cite{Bernevig2012}. To come up with an elegant formal classification, the spontaneous breaking of inversion symmetry by the CDW has to be carefully accounted for. We postpone this issue to a forthcoming publication, where related models will be discussed \footnote{{H.H. Jen, M. H\"oning, F. Grusdt and M. Fleischhauer, in preparation.}}. Here we restrict ourselves to the definition and calculation of a topological invariant $\nu$, which is quantized to $\nu=0,\pi$ and protected by inversion symmetry. 

The topological invariant we employ is the many-body Zak or Berry phase defined by twisted boundary conditions along the ladder \cite{Resta1995,Grusdt2013EdgeStates}. Like in the case of the Chern number in the $1+1$D case, we introduce the twist angle $\theta_y$, however now the second parameter $\theta_x = \pm \pi/2$ is fixed. In practice the most convenient way to implement twisted boundary conditions is to multiply the hopping elements from the last to the first sites of the ladder (which realize periodic boundary conditions) by the complex phase $e^{i \theta_y}$. Then the eigenstate $\ket{\Psi(\theta_y)}$ depends on $\theta_y$ and the Berry phase can be calculated as usual,
\begin{equation}
\nu = \int_0^{2 \pi} d \theta_y ~ \bra{\Psi(\theta_y)} i \partial_{\theta_y} \ket{\Psi(\theta_y)}.
\label{eq:defBerry}
\end{equation}
From inversion symmetry it follows that $\nu=0,\pi$ is strictly quantized \cite{Zak1989,Hatsugai2006}.

To calculate the topological invariant $\nu$, we restrict ourselves to the simple representation \eqref{eq:CDWgutzwiller} of the CDW state $\ket{\Psi}$. Then we distinguish four different cases, characterized by $\theta_x = \pm \pi / 2$ and by which of the two states $\ket{\text{CDW}}$ and $\ket{\text{CDW}'}$ we use. To begin with we note that only for, say, $\theta_x=\pi/2$ the link with the complex phase $e^{i \theta_y}$ is part of an atomic orbital, as defined in the discussion of FIG.\ref{fig:ThoulessPumpOrbitals}. Then in the trivial case $\theta_x=-\pi/2$, $\ket{\Psi}$ is independent of $\theta_y$ and thus $\nu=0$ vanishes for both CDW states. For $\theta_x=+\pi/2$ on the other hand, we have to distinguish between CDW and CDW'. Only for one of the two states -- say for $\ket{\text{CDW}}$ -- the link with the complex phase $e^{i \theta_y}$ is part of an \emph{occupied} atomic orbital. Thus for the state described by CDW' the wavefunction $\ket{\Psi}$ is independent of $\theta_y$ and $\nu=0$ again. Finally we will show that the state CDW is topologically non-trivial with $\nu=\pi$. To this end, note that there is an occupied atomic orbital on the link connecting the last and the first rung of the ladder. The energy of this orbital can not be changed by the complex phase $e^{i \theta_y}$, which is merely a gauge transformation, but the eigenfunction of the orbital $\psi_m(\theta_y)$ (with $m=1,...,4$ labeling the four sites), depends on $\theta_y$. In fact, a simple calculation shows that the corresponding Berry phase is $\int_0^{2 \pi} d \theta_y \sum_m \psi_m^* i \partial_{\theta_y} \psi_m = \pi$. Because $\ket{\text{CDW}}$ is a simple product state it follows that $\nu=\pi$ in this case.

%%%%%%%%%%%%%%%%%%%%%%%%%%%%%%%%%%%%%%%%%%%%%%%%%%%%%
\section{Summary and Outlook}
\label{sec:OutlookSummary}
%%%%%%%%%%%%%%%%%%%%%%%%%%%%%%%%%%%%%%%%%%%%%%%%%%%%%
In summary, we have proposed and analyzed a realistic setup for the realization of a topologically non-trivial CDW state (at filling $\rho=1/8$) of strongly interacting bosons in a 1D ladder geometry. Our model was derived by taking the thin-torus limit of the 2D Hofstadter-Hubbard model at flux $\alpha=1/4$ per plaquette. The $\nu=1/2$ Laughlin-type fractional Chern insulator in this 2D model is directly related to the 1D CDW at filling $\rho=1/8$. As a consequence, the CDW has interesting topological properties: When adiabatically introducing magnetic flux $\theta_x/2 \pi$ through the small perimeter of the thin torus, which can be realized by changing the hoppings in our model, a fractionally quantized Hall current is induced along the ladder. Alternatively, the CDW phase can be interpreted as inversion symmetry-protected topological phase, characterized by a quantized topological invariant taking values $\nu=0,\pi$. We used DMRG calculations to determine the particle-hole gap of the CDW and found values of $\Delta_\text{CDW} \sim 0.1 J$, a sizable fraction of the bare hopping $J$. When placed in a harmonic trap, the wedding cake structure of the density provides a clear signature of the appearance of the topological CDW state.

Investigating the thin-torus limit of fractional Chern insulators is a promising route to gain understanding of more complicated, but closely related, topologically ordered states in 2D systems \cite{Bergholtz2005,Bergholtz2006,Bernevig2012,Papic2014}. In this work we showed how the thin-torus limit can be realized experimentally with ultracold atoms, including the possibility of fully tunable twisted boundary conditions. Similar ideas can be carried over to photonic systems, where synthetic gauge fields can also be implemented \cite{Otterbach2010,Hafezi2011,Umucalilar2011,Hafezi2013a} and strong non-linearities on a single-photon level are realized e.g. using Rydberg atoms \cite{Lukin2001,Peyronel2012}. Therefore an interesting future direction for such experiments would be the observation of more complicated thin-torus models, going beyond the analogue of the simple $1/2$ LN state and including for instance states related to the non-Abelian Read-Rezayi series \cite{Read1999,Bergholtz2006,Papic2014}. 

Once a system like the one described in this paper is realized, an important question is how to witness its topological properties. The quantized transport connected to the Chern number could be measured by taking in-situ images of the atomic cloud. A more direct measurement of the topological invariant would be desirable, which should also be able to measure the invariant $\nu$ characterizing the symmetry-protected topological order. Such measurements have been performed in non-interacting systems \cite{Atala2012,Duca2014} using a combination of Ramsey-interferometry and Bloch oscillations, and they could be extended to interacting systems in the future \footnote{F. Grusdt, N. Yao, D. Abanin and E. Demler, in preparation.}.

%%%%%%%%%%%%%%%%%%%%%%%%%%%%%%%%%%%%%%%%%%%%%%%%%%%%%
%%%%%%%%%%%%%%%%%%%%%%%%%%%%%%%%%%%%%%%%%%%%%%%%%%%%%
\section*{Acknowledgements}
%%%%%%%%%%%%%%%%%%%%%%%%%%%%%%%%%%%%%%%%%%%%%%%%%%%%%
%%%%%%%%%%%%%%%%%%%%%%%%%%%%%%%%%%%%%%%%%%%%%%%%%%%%%
The authors thank M. Fleischhauer for supporting this work. They would also like to thank M. Fleischhauer, M. Atala, M. Aidelsburger, M. Lohse and C. Schweizer for fruitful discussions. F.G. was supported by a fellowship through the Excellence Initiative (DFG/GSC 266) and he gratefully acknowledges financial support from the "Marion K\"oser Stiftung". The financial support of SFB/TR 49 is gratefully acknowledged.

%\bibliography{/Users/fgrusdt/Documents/PhD/JabRef/dataBase_JabRef2.bib}

\end{document}